# Transfer of graphene onto arbitrary substrates via sublimable carrier

Yu-Hao Deng[1*]

[1] Academy for Advanced Interdisciplinary Studies, Peking University, Beijing, China

* Correspondence should be addressed to yuhaodeng@pku.edu.cn

**Abstract**

Graphene, a monolayer of carbon atoms packed into a two-dimensional crystal structure, attracted intense attention owing to its unique structure and optical, electronic properties. Recent advances in chemical vapor deposition (CVD) have led to the batch production of high quality graphene on metal foils. However, further applications are required in the way these graphenes are transferred from their growth substrates onto the target substrate. Here, we report a sublimable carrier method that allows the graphene to be transferred with high quality onto arbitrary substrates, including semiconductor, metal and organic substrates. The intrinsic problems of the residue and environmentally unfriendly organic solvents have been solved due to the polymer-free process. Optical microscopy, scanning electron microscopy (SEM) and Raman spectroscopy demonstrate the high quality and clean surface of the transferred graphene. This work provides a new way of optimizing graphene transfer and widens the applications of graphene in large-scale 2D electronics.

**Introduction**

Graphene has demonstrated potential for future semiconductor and photonic technologies due to its remarkable electrical and optical properties, including an unusual quantum Hall effect combined with high carrier mobility, excellent optical conductivity, and linear dispersion of the Dirac electrons [1-5]. Chemical vapor deposition (CVD) has been proved to be the most suitable method for batch production of high quality graphene. For engineering applications, such as solar cells, transistors, photodetectors, touch screens, and flexible smart windows, a subsequent transfer process of graphene from metal substrates (Cu or Ni) [6, 7] to target substrates is necessary.

Thus far, the PMMA-carrier method is most commonly used for transferring graphene from metal substrates to arbitrary substrates [8-10]. In this method, polymethyl methacrylate (PMMA) is spin-coated onto graphene as carrier material, and then the metal substrate is etched away by etching reagents. Subsequently, the PMMA/graphene film is transferred onto arbitrary substrates. Finally, The PMMA is dissolved using organic solvents, leaving the graphene layer on the target substrates. Although PMMA-carrier method is a straightforward transfer process, the organic solvents treatment frequently fails to fully remove the PMMA, leaving the residue on the graphene surface and substrates [11]. Meanwhile, large amounts of organic solvents used to remove PMMA layer is not environmentally friendly. Several improved strategies have been proposed to remove the residue, including change organic solvents and polymers [12], using modified RCA cleaning processes [11], mechanically sweeping by probe [13], or further surface cleaning [14]. However, these techniques or other complicated wet chemistry methods are limited to completely remove the residue and be environmentally friendly due to the intrinsic use of polymers and organic solvents.

Herein, we report a polymer-free method for graphene transfer. By introducing sublimable material instead of polymer layer as the carrier, the intrinsic problems of the residue and organic solvents have been solved. Optical microscopy, scanning electron microscopy (SEM) and Raman spectroscopy are used to characterize the transferred graphene. This new transfer method can also be applied in transferring graphene to arbitrary substrates.

## Results and discussion

Fig. 1 shows the schematic illustration of this new transfer processes. Sublimable naphthalene was evaporated and deposited on the graphene surface, then the growth substrate (Cu foil) was etched using FeCl$_3$ solution (1M FeCl$_3$, H$_2$O). Subsequently, the naphthalene/graphene was rinsed by deionized water to remove residual Fe$^{3+}$ ions and transferred to the target substrate. Finally, put the stack in vacuum environment or mild temperature (~ 50 °C) to sublimate the naphthalene layer, graphene was successfully transferred onto the target substrate.

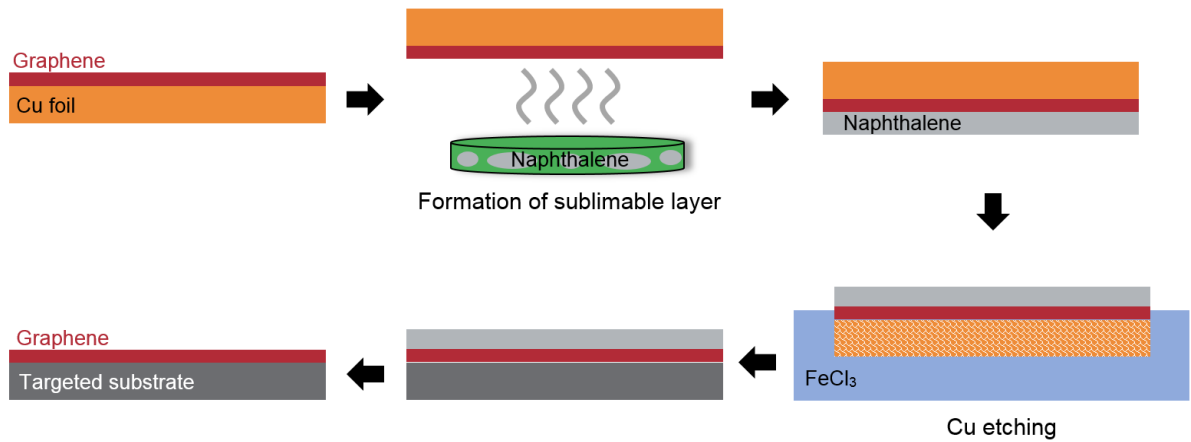

**Fig. 1** Schematic illustration of graphene transfer via sublimable carrier. Sublimable naphthalene was deposited on the graphene surface as the carrier, then the growth substrate was removed using etching solution. Subsequently, the naphthalene/graphene was transferred to the target substrate. Finally, sublimating naphthalene in vacuum environment or mild temperature, the graphene was successfully transferred onto the target substrate.

Due to no polymer carrier was used in the new method, there will be any residue remains on the graphene surface theoretically. There also will be no organic solvents to remove the polymer, which contributes this approach to environmentally friendly. As shown in Fig. 2A, residues on the surface is visible in the optical microscopy image of graphene on $SiO_2$/Si substrate transferred by PMMA carrier method. The polymer residue appears as rounded particles under SEM (Fig. 2B). The PMMA residues on graphene would affect the quality of graphene [15]. The Raman spectroscopy is a widely used and unambiguous technology to evaluate the number of layers and the quality of graphene [16, 17]. The difference of the D, G and 2D peak indicates the quality and uniformity of the graphene. Fig. 2C shows the Raman spectra of monolayer graphene transferred by PMMA carrier method. Differing from the polymer carrier method, the sublimable material will not leave any residue here. So, we got a cleaner graphene surface shows in Fig. 2D by naphthalene carrier method. The SEM image also confirmed that there is no significant residue on the graphene surface the result (Fig. 2E). Fig. 2F shows the Raman spectra of monolayer graphene transferred by naphthalene carrier method. The occurrence of D peak in PMMA transferred graphene is due to the residues or wrinkles formed during the transferred process [18, 19]. However, the D peak of the naphthalene carrier method transferred graphene is absent, suggesting that the graphene is well-transferred on the wafer without obvious residues and damage. The increase of the 2D/G ratio of graphene transferred by naphthalene carrier method also indicates a cleaner surface with lower doping level [20]. By introducing sublimable material instead of polymer layer as the carrier, our polymer-free transfer method obtained a cleaner graphene surface than conventional method.

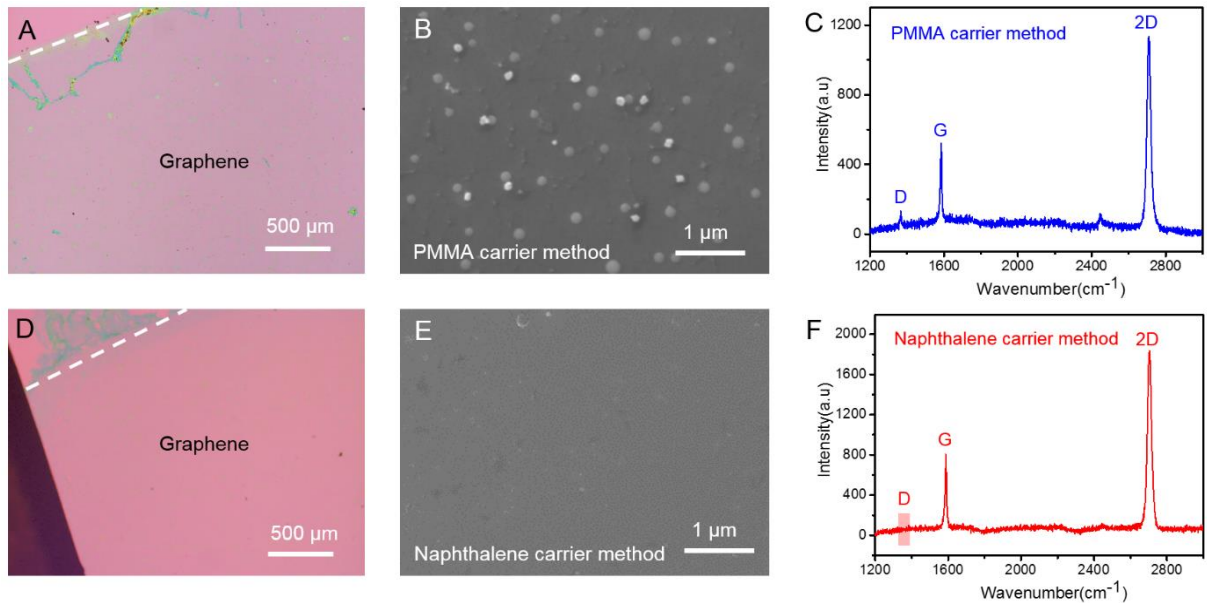

**Fig. 2** Comparison of PMMA carrier and naphthalene carrier method. Optical microscopy (A), scanning electron microscopy (SEM) (B) and Raman spectroscopy (C) of graphene on $SiO_2$/Si substrate transferred by PMMA carrier method. Optical microscopy (D), scanning electron microscopy (SEM) (E) and Raman spectroscopy (F) of graphene on $SiO_2$/Si substrate transferred by naphthalene carrier method.

For various engineering applications, the transfer process of graphene from metal substrates to arbitrary substrates is necessary. Based on naphthalene carrier method, we have successfully transferred graphene to $SiO_2$/Si, metallic Au, and polyethylene terephthalate (PET) flexible substrate (Fig. 3A-C). Particularly, benefit from the absence of organic solvents during the overall process, this naphthalene carrier method is compatible with organic substrates for flexible electronic applications.

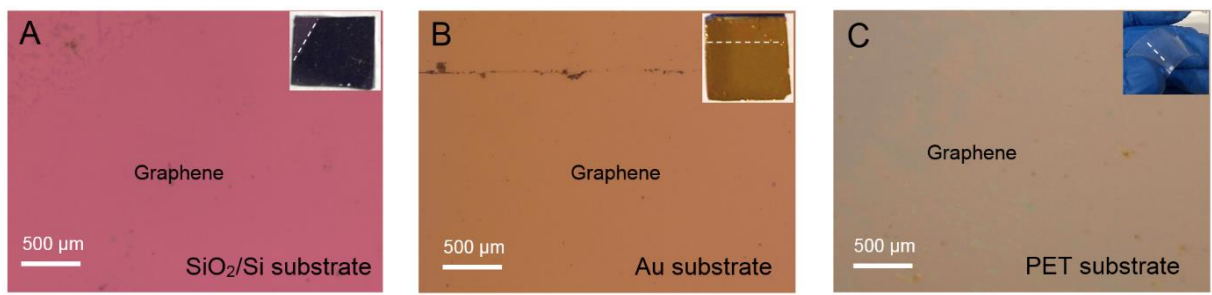

**Fig. 3** Transfer of graphene onto arbitrary substrates. graphene on $SiO_2$/Si (A), Au (B) and PET (C) substrate transferred by naphthalene carrier method.

## Conclusions

Here we present a polymer-free method to transfer high-quality graphene from growth substrate onto arbitrary substrates. By using sublimable naphthalene instead of the traditional polymer as the carrier layer, the long-standing problems of residue and environmental protection have been solved. In addition to the field of graphene transfer, other two-dimensional materials and nano materials can also be transferred via this method. This work provides a new way of optimizing graphene transfer and will be useful for applications in large-scale 2D electronics.

## Methods

**Graphene growth:** Graphene was grown on Cu foil (99.8%, 25 mm thick, Alfar Aesar) via low pressure CVD. Cu foil was placed on a quartz boat and inserted in a quartz tube inside a tube furnace. The system was heated to 1000 ºC in 60 minutes with $H_2$ (20 sccm) at a pressure of 50 Pa, and kept at 1000 ºC for another 30 minutes to anneal the Cu foil. After that, $CH_4$ (20 sccm) was introduced with a total pressure of 100 Pa for 30 minutes. Finally, thesy stem was naturally cooled down to room temperature under the mixture of $H_2$ and $CH_4$ ambient.

**PMMA carrier method:** A layer of PMMA (950 K, 4%) was spun on the graphene grown on Cu foil. Then, the Cu foil was etched off by 1 M $FeCl_3$ for 30 minutes. After being rinsed by deionized water for several times, the PMMA/graphene was transferred onto the $SiO_2$/Si substrate. Finally, the PMMA was removed by acetone.

**Characterization:** Optical images were acquired using optical microscope (Zeiss Axio Imager, A2m) and Raman spectra were acquired using Raman spectroscopy (JY LabRam HR800). SEM images were obtained using a Helios NanoLab DualBeam electron microscope, an accelerating voltage 5 kV and beam current of 87 pA was used for all samples.

## Data availability

All data are available from the corresponding author(s) upon reasonable request.

**Acknowledgements:** None.

**Conflict of interest:** The authors declare no competing financial interest.

**Note added:** This work was patented in 2015 (CN, ZL 201510962569.0, A method of the transfer graphene based on sublimed method, 2015, Peking university).